\documentclass[fleqn,twoside]{article}
\usepackage{espcrc2}
\usepackage{graphicx}
\usepackage[figuresright]{rotating}

\newcommand{\AmS}{{\protect\the\textfont2
  A\kern-.1667em\lower.5ex\hbox{M}\kern-.125emS}}
\newcommand{\lsim}{\mathrel{\mathop{\kern 0pt \rlap
  {\raise.2ex\hbox{$<$}}}
  \lower.9ex\hbox{\kern-.190em $\sim$}}}

\newcommand{\gsim}{\mathrel{\mathop{\kern 0pt \rlap
  {\raise.2ex\hbox{$>$}}}
  \lower.9ex\hbox{\kern-.190em $\sim$}}}

\title{Candidates for non--baryonic dark matter}

\author{Nicolao Fornengo\address{Dipartimento di Fisica Teorica, Universit\`a di Torino
    and INFN, Sezione di Torino \\ via P. Giuria 1, I--10125 Torino, Italy}}
       
\begin{document}
\begin{abstract}
  This report is a brief review of the efforts to explain the nature of
  non--baryonic dark matter and of the studies devoted to the search for relic
  particles. Among the different dark matter candidates, special attention is
  devoted to relic neutralinos, by giving an overview of the recent
  calculations of its relic abundance and detection rates in a wide variety of
  supersymmetric schemes.  \vspace{1pc}

%  This report is a brief review of the efforts to explain the observed amount
%  of non--baryonic dark matter and of the studies of the possible signals which
%  can be looked at for trying to detect the relic particles which compose the
%  dark matter component of the Universe. Among the different dark matter
%  candidates, special attention is devoted to relic neutralinos, by giving an
%  overview of the recent calculations of its relic abundance and detection
%  rates in a wide variety of supersymmetric schemes.  \vspace{1pc}
\end{abstract}

% typeset front matter (including abstract)
\maketitle

\section{Evidence of darkness}

The presence of large amounts of non--luminous components in the Universe has
been identified along the years by different means and on different scales: on
the galactic scale, the flatness of the rotational curves of many galaxies
indicates a dark component which is presumably distributed as a halo around the
galaxies \cite{Rot.curves}; clusters points toward a sizeable contribution of
unseen matter distributed around and between galaxies \cite{Clusters}; more
recently, on cosmological scales, the combination of the results on
high--redshift supernovae \cite{Highz.SN} and on the anisotropies of the cosmic
microwave background radiation \cite{CMBR} is pointing toward a flat Universe
whose energy density is dominated by a dark vacuum component (cosmological
constant, quintessence) together with a sizeable dark component of matter. In
terms of the density parameter $\Omega$, the current view can be summarized as
follows \cite{deBernardis}: the total amount of matter/energy of the Universe
is $\Omega_{\rm tot} = 1.02 \pm^{0.06}_{0.05}$, and this is composed of a
matter component $\Omega_{\rm M} = 0.31 \pm^{0.13}_{0.12}$ and a vacuum--energy
component $\Omega_{\Lambda} = 0.71 \pm {0.11}$. Even though the actual numbers
vary a little depending on the priors of the statistical analyses, the clear
indication of the latest data is that the Universe is strongly dominated by
dark (and unknown) components. In fact the numbers above cannot be reconciled
with a Universe made only of standard components: from primordial
nucleosynthesis studies, baryons can contribute only at the level of
$\Omega_{\rm b} = 0.037 \pm 0.011$ \cite{Baryons}, while luminous matter is
known to provide only a contribution of order $\Omega_{\rm lum} \sim 0.003$
\cite{Luminous}. We are therefore facing the presence of at least {\em three
  dark components} in the Universe: dark baryons, dark non--baryonic matter and
dark energy.  The existence of both dark (relativistic or non--relativistic)
exotic matter and dark energy asks for extension of the standard model of
fundamental interactions, since no known particle or field can explain either
of these components.

In this review, we will discuss the current status of the non--baryonic dark
matter problem. For reviews on dark baryons and dark energy, see Refs.
\cite{Talk.b,Talk.l}. We will discuss the efforts to explain the amount of dark
matter in the Universe, which we summarize as:
\begin{equation}
0.05 \lsim \Omega_{\rm M} h^2 \lsim 0.3
\label{eq:DM}
\end{equation}
and the studies related to the searches for dark matter particles.

\vspace{-5pt}
\section{Particle candidates to dark matter}

Dark matter candidates must be looked for in schemes beyond the
standard model. The interesting feature of many of these models is
that they are extensions whose primary motivation is not related to
dark matter: in spite of this, they contain particles which have the
right properties to act as dark matter. This is especially true for
supersymmetric theories, which are the ones with richer phenomenology
for dark matter, but interesting candidates may also be found in
theories which do not involve supersymmetry.

Among the non--supersymmetric candidates, the {\em massive neutrino} and
the {\em axion} are the most direct possibilities.

A light massive neutrino represents the simplest extension of the standard
model. The recent results on the atmospheric \cite{Nu.atm,Talk.nu} and solar
\cite{Talk.nu,Nu.solar} anomalies strongly indicate that neutrinos do possess a
mass: a consequence of this is that neutrinos could also compose at least a
fraction of dark matter. From the atmospheric neutrino analyses \cite{Nu.atm},
we deduce that neutrinos can provide $\Omega_{\nu} h^2 \gsim 6 \cdot 10^{-4}$,
but very likely they cannot contribute more than $\Omega_{\nu} h^2 \sim 0.05$
in order not to spoil the process of structure formation. Therefore, a massive
neutrino can only be (and very likely is) present as a sub--dominant dark
matter component. Although less economic from the point of view of particle
physics, it is also possible to have very massive neutrinos ($m_\nu \gsim 100$
GeV) which again would contribute only partially, at the level of $\Omega_{\nu}
h^2 \sim 10^{-3}$--$10^{-2}$ \cite{Enqvist}. Sterile neutrinos with masses in
the keV range and mixing angles of the order of $10^{-10}$ rad may play a role,
since they may provide $\Omega_{\nu} h^2 \sim 0.1$ -- $0.2$
\cite{Dolgov.Fuller}, and therefore explain the dark matter amount of
Eq.(\ref{eq:DM}).

Axions arise in models where the strong CP problem is solved by the
Peccei--Quinn mechanism. They may be produced in the early Universe by
means of different mechanisms, either thermally or non--thermally
(misalignment of the axionic field, decay of axionic strings)
\cite{Axion}. The current limits \cite{Scopel.axion} from axion searches and
cosmology constrain this particle to have either a mass around 10 eV
or in the mass range $10^{-6}$--$10^{-2}$ eV. In both cases, axions
can solve the dark matter problem stated in Eq.(\ref{eq:DM}), but they
can also be a sub--dominant component.

Supersymmetry with conserved R--parity offers a wide variety of dark matter
candidates. The nature of the relic particle depends somehow on the way
supersymmetry is broken. Gravity--mediated models predict as dark matter
particle either the {\em neutralino} or the {\em sneutrino} \cite{Sneutrino}.
The neutralino may be the dark matter candidate also in anomaly--mediated
models. In gauge mediated--models the role of dark matter may be played by the
{\em gravitino} \cite{Gravitino} or by {\em messenger fields}
\cite{Messenger.fields}. In supersymmetric models where the strong CP problem
is solved by the axion, its supersymmetric partner, the {\em axino}
\cite{Axino} is also present and can act as dark matter. Other possibilities
that have been proposed are stable non--topological solitons, like the {\em
  Q--balls} \cite{Q.balls}, or {\em heavy non--thermal relics} \cite{Zillas}.
The most interesting candidate, and the one which has been more deeply
investigated, is the neutralino, and in the following Sections we will
concentrate our discussion on this particle. For the other supersymmetric
candidates, we refer to the quoted references and the references quoted
therein.

Finally, among other candidates which have been proposed, not necessarily
related to supersymmetry, we finally recall {\em mirror particles}
\cite{Mirror} or some type of {\em scalar fields} \cite{Scalar.fields}.

From the above discussion on particle dark matter candidates, a fact clearly
arises: non--baryonic dark matter may naturally be multi--component. An example
of this occurs, for instance, in models where the CP problem is solved by the
axion and the hierarchy problem is explained by low--energy supersymmetry. In
this case, the axion and the axino or the neutralino are all dark matter
components, plus the massive neutrino which nowadays appears to be an
unavoidable dark matter component. Which one would turn out to be the dominant
candidate would depend on the characteristics of the specific model which one
can build. But the possibility of multi--component dark matter is interesting,
because the detectability of a specific candidate is not in general related to
the fact that this particle is the dominant component of dark matter. Usually,
at least for candidates like the neutralino in standard cosmology, it turns out
that it is easier to detect a relic particle which is sub--dominant
\cite{Bottino:2001it}.  This is related to the fact that detection rates rely
on the cross sections of the processes of scattering or annihilation, while on
the contrary the relic abundance is inversely proportional to the annihilation
cross section. This roughly induces an anti--correlation between detection
rates and cosmological abundance: therefore large detection rates are often
related to low relic abundances. It is not trivial to have a relic particle
which is at the same time a dominant dark matter component and potentially
detectable. Clearly this last situation is the most appealing one.

As for the most studied candidate which is the neutralino, in the
following it will be shown that this situation can occur in suitable
classes of supersymmetric models. This is one of the properties that
makes the neutralino a very appealing candidate.

\vspace{-5pt}
\section{Supersymmetry and dark matter: the neutralino}

The existence of a relic particle in supersymmetric theories arises from the
conservation of a symmetry, $R$--parity, which prevents the lightest of all the
superpartners from decaying. The nature and the properties of this particle
depend on the way supersymmetry is broken, and we have already seen that the
neutralino can be the dark matter candidate in models where supersymmetry is
broken through gravity-- (or anomaly--) mediated mechanisms. The actual
implementation of a specific supersymmetric scheme depends on a number of
assumptions on the structure of the model and on the relations among its
parameters. This induces a large variability of the phenomenology of
neutralino dark matter. In this Section we briefly recall the supersymmetric
models which have been mostly studied for neutralino dark matter, while in the
next Sections we will give a few examples of calculations of relic abundance and
detection rates in some of these schemes. For details concerning the models and
their implementations for neutralino dark matter calculations, we refer to the
list of references provided below and references quoted therein.

The simplest and most direct implementation of supersymmetry is represented by
the {\em minimal supergravity} (mSUGRA) scheme
\cite{Berezinsky:1995cj,Berezinsky:1996ga,Bottino:1998ka,Bottino:2000jx,Bottino:2000kq,Ellis.sugra-u.omega,Omega.fig,Ellis.direct-u.omega,Focus-point,Arnowitt.sugra.direct,Roszkowski.sugra.relic,Vergados.sugra.direct,Drees.sugra.direct,Mandic.direct,Lahanas.sugra.direct,Upmu.sugra.u-nu,Jeannerot:1999yn},
where, in addition to requiring gauge coupling constant unification at the GUT
scale, also all the mass parameters in the supersymmetric breaking sector are
universal at the same GUT scale. The low--energy sector of the model is
obtained by evolving all the parameters through renormalization group equations
(RGE): this process also induces the breaking of the electroweak (EW) symmetry
in a radiative way (rEWSB). This model is very predictive, since it relies only
on four free parameters, but at the same time it induces a very constrained
phenomenology at low--energy. It also appears to be quite sensitive to some
standard model parameters, like the mass of the top and bottom quarks ($m_t$
and $m_b$) and the strong coupling constant $\alpha_s$.

A more relaxed implementation of this supersymmetric scheme is offered by {\em
  non--universal supergravity} (nuSUGRA)
\cite{Berezinsky:1995cj,Berezinsky:1996ga,Bottino:1998ka,Bottino:2000jx,Bottino:2000kq,Arnowitt.sugra.direct,Upmu.sugra.u-nu,Ellis.direct-nu.omega,Gaugino-nu,Drees.sugra-nu.omega},
where some of the unification conditions at the GUT scale are relaxed:
non--universality has been studied in the Higgs, in the sfermion and in the
gaugino sectors.

Specific patterns of non--universality may be originated through mechanisms
which involve effects of extra--dimensions, like in {\em D--brane}
\cite{Arnowitt.brane.omega,Arnowitt.brane.direct,Nath.brane.direct} and {\em
  string models} \cite{Nath.brane.direct}. These models also may be very
predictive, with very few free parameters, but the relations among them is
different from what is postulated in mSUGRA models.

It has also been realized that unification conditions, both for gauge couplings
and/or mass parameters, may occur at scales which are different from the
standard GUT scale at about $10^{16}$ GeV. This unification scale
may be lower than the usual GUT scale ({\em intermediate unification scale
  models}) \cite{Intermediate-scale,Intermediate-scale.fig}, and this induces a
modification of neutralino phenomenology at low energy.

A different approach is offered by the {\em low--energy supersymmetric model}
(effMSSM)
\cite{Bottino:2001it,Bottino:2000jx,Bottino:2000kq,Mandic.direct,Bottino:2000gc,Bottino:1999ei,Belli:1999nz,Bottino:1998vw,Bottino:1998tw,Bottino:1998jx,Bottino:1994xp,Gondolo.mssm.omega,Klapdor.mssm.direct,Galactic.center,Gamma,Gamma.fig,Bergstrom.mssm.upmu,Clumpy,Positrons.HEAT,Dbar,Antiproton,Extra_G:gamma,Extra_G:direct},
defined directly at the EW scale, which is where the phenomenology of
neutralino dark matter is actually studied. Also in this case we have to make
assumptions in order to reduce the number of free parameters to a manageable
number. These assumptions must be mild enough not to represent an arbitrary
over--constraint on the model, and all the relevant parameters at the EW scale
must be represented. It is possible in this case to work with six or seven free
parameters.

Other models which have been discussed in the literature in connection with
neutralino dark matter are {\em dilaton domination} models \cite{Dilaton},
models with {\em CP--violation} \cite{CP-violation} and {\em anomaly mediated
  models} \cite{Anomaly.mediated}.

%%%
The parameter space of all the models is constrained by a number of relevant
experimental bounds which mostly come from accelerator physics. Negative
searches of supersymmetry at LEP \cite{LEP} and Tevatron \cite{Tevatron} are
currently constraining charginos to be heavier than about 105 GeV and sleptons
and squarks to be heavier than 90--100 GeV. The bounds from LEP on the lightest
supersymmetric Higgs boson $h$ indicate that its mass is larger than 90 GeV for
small coupling of $h$ with $b$--quarks, while for intermediate and large values
of this coupling the bound on the mass is about 114 GeV \cite{LEP}. Tevatron
further constrains the mass of $h$ in some sectors of the supersymmetric
parameter space (namely, large values of $\tan\beta$, the ratio of the vev's of
the two neutral higgses) \cite{Tevatron}. Finally, an important constraint for
all supersymmetric models comes from the measured values of the branching ratio
of the radiative decay $b\rightarrow s + \gamma$, which, in addition to the
standard model terms, can get sizeable contributions from supersymmetric loops
\cite{bsg}. Notice that the requirement that the neutralino is a dark matter
particle, implies that also cosmology imposes a limit on supersymmetric models:
in fact, the relic abundance of neutralinos, which can be calculated in a given
supersymmetric scheme, cannot exceed the upper limit of Eq.(\ref{eq:DM}).

At present, there is no direct indication that supersymmetry has been found in
Nature. However, we just want to mention a few recent results which have been
considered as possible hints of supersymmetry. The first is the excess of
signal--like events in the Higgs searches at LEP: an effect of about
2--$\sigma$ for a Higgs mass around 115 GeV has been reported \cite{LEP.Higgs}.
This light higgs may be interpreted as indicative of a supersymmetric higgs.
Another indication for physics beyond the standard model comes from a global
fit to the precision electroweak data: the analysis shows that the standard
model is acceptable with a low (around 4\%) confidence level \cite{Altarelli}.
The agreement with the data improves considerably if low--energy supersymmetry
is included in the analysis \cite{Altarelli}.

In the last year another anomaly raised a lot of attention: a possible
deviation between the calculated and the measured value of the anomalous
magnetic moment of the muon. However, recent theoretical re--evaluations have
shown that the evidence of deviation is very weak \cite{muong2}, and therefore
instead of indicating a possible presence of supersymmetry, the results on the
muon anomalous magnetic moment should be regarded as imposing bounds on the
supersymmetric parameter space, as is the case for $b\rightarrow s + \gamma$.

\section{Neutralino relic abundance}

\begin{figure}[t]
\vspace{4pt}
\includegraphics*[width=0.45\textwidth]{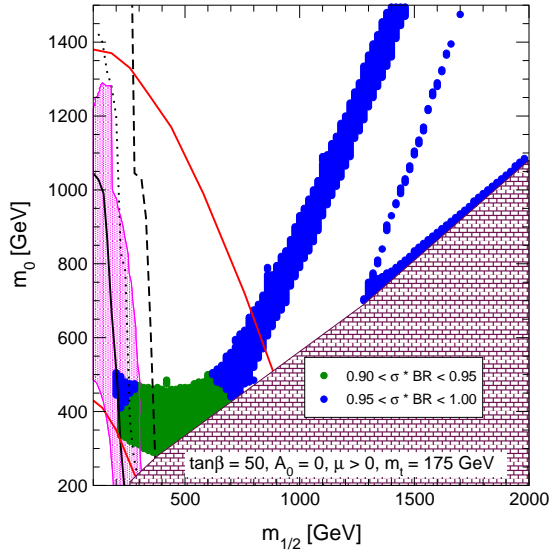}
\vspace{-30pt}
\caption{Neutralino relic abundance in a mSUGRA scheme, for
  $\tan\beta=50$ and $m_t=175$ GeV \cite{Omega.fig}. Inside the dark (green and
  blue) bands: $0.1 \leq \Omega_\chi h^2 \leq 0.3$.}
\label{relic}
\end{figure}
Neutralino relic abundance, like also neutralino detection rates, is sensitive
to the specific supersymmetric model which is employed. In fig. \ref{relic} we
show an example, taken from Ref. \cite{Omega.fig}, which refers to a mSUGRA
scheme
\cite{Berezinsky:1995cj,Bottino:1998ka,Bottino:2000kq,Ellis.sugra-u.omega,Omega.fig,Arnowitt.brane.omega,Belanger:2001fz}
in the large $\tan\beta$ regime. The regions of the parameter space shown in
the figure as dark (blue and green) bands entail a relic neutralino with relic
abundance in the range $0.1 \leq \Omega_\chi h^2 \leq 0.3$.  Therefore, in that
sector of the parameter space the neutralino represents the dominant dark
matter component since it is able to fulfill the cosmological requirement of
Eq.(\ref{eq:DM}). Instead, in the region which is between the thick and the
thin (blue) bands, neutralino relic abundance is smaller than 0.1, and in a
part of this region it falls below the lower limit of Eq.(\ref{eq:DM}): in this
last case, neutralino cannot account for all the dark matter in the Universe.
On the contrary, in the other two large white regions of fig. 1, $\Omega_\chi
h^2 > 0.3$: these regions of the mSUGRA parameter space have to be considered
as excluded by the requirement that neutralino relic abundance does not get in
conflict with the experimental determination of the amount of dark matter in
the Universe. In the same parameter space of fig. 1, another region has been
identified where $\Omega_\chi h^2$ can fall in the range of Eq.(\ref{eq:DM}):
this can occur at large $m_0/m_{1/2}$ \cite{Berezinsky:1995cj,Focus-point}, a
region which has been also called as ``focus point model'' ($m_0$ is the common
soft scalar mass parameter at the GUT scale and $m_{1/2}$ is the common gaugino
mass at the GUT scale).

The situation shown if Fig.1 can change substantially if one explores other
sectors of the mSUGRA parameter space, like for instance small values of
$\tan\beta$, where the region in which neutralino is the dominant component is
quite constrained to values of $m_0$ and $m_{1/2}$ which do not exceed a few
hundreds of GeV, except in very narrow corridors where coannihilation between
neutralinos and staus occur \cite{Omega.fig}.

It has to be noticed that mSUGRA models turn out to be somewhat sensitive to
the standard model parameters $m_t$, $m_b$ and $\alpha_s$ and to the details of
radiative effects (radiative corrections, radiative electroweak symmetry
breaking). It is therefore difficult to derive definite conclusions on which
are the intervals of the mSUGRA parameters where neutralino is a dominant dark
matter component (except for small values of $\tan\beta$).

The effects of non--universality
\cite{Berezinsky:1995cj,Bottino:1998ka,Bottino:2000kq,Ellis.direct-nu.omega,Drees.sugra-nu.omega,Arnowitt.brane.omega}
affect the low--energy sector of the theory through changes in the RGE and in
the occurrence of rEWSB. This has the effect of modifying neutralino couplings
and the mass spectrum of supersymmetric particles, with the consequence of
modifying neutralino relic abundance and the cosmologically relevant regions in
the nuSUGRA parameter space. Quite relaxed relic abundance regions are also
obtained in the low--energy realization of supersymmetry, the effMSSM
\cite{Bottino:2001it,Bottino:2000kq,Mandic.direct,Bottino:1998jx,Gondolo.mssm.omega,Gamma}.

\section{Relic neutralino searches}
Neutralino relic particles can be searched for by means of different
techniques. Basically, the detection methods are of two types: {\em
direct detection}, which relies on the possibility to directly measure
the interaction of neutralinos with a detector in the laboratory, or
{\em indirect detection}, which tries to identify products of
neutralino annihilation.

Dark matter particles are supposed to be bounded to the Galaxy, since
they are part of the dark halo which is responsible for the observed
flatness of the rotational curves. Therefore the best place to look
for relic neutralinos is the halo of our Galaxy. The dynamics and
distribution of these particles are described by a matter-- and a
velocity-- distribution function, about which there are still large
uncertainties. 

The distribution in space of these bounded dark particles can be smooth and
non--singular at the galactic center, but numerical simulations tend to
indicate a cuspy behaviour toward the center of the Galaxy as well as a clumpy
distribution of dark matter in the halo. Detection rates which depend on
annihilation processes which occur in the galactic halo are therefore strongly
affected by this uncertainty in the matter distribution. On the contrary,
detection rates which rely on local properties, like direct detection, are not
very much sensitive on how dark matter is distributed in the halo. In this
case, the relevant parameter is the total local dark matter density $\rho_l =
(0.2 \div 0.7)$ GeV cm$^{-3}$ \cite{Bottino:2000jx}. Neutralinos can account
for all or part of this amount, depending whether they are or not the dominant
dark matter component.  Therefore, the local neutralino matter density is:
$\rho_\chi = \xi \rho_l$, where $\xi \leq 1$ \cite{Bottino:2001it}.

The velocity distribution of dark matter particles in the halo is even more
uncertain than the matter distribution, since it is difficult to constrain the
way these dark particles move in the halo. The usual assumption is to consider
an isotropic Maxwell--Boltzmann (MB) distribution, as seen in the galactic rest
frame. The relevant parameters in this case are only two: the local rotational
velocity $v_0 = (220 \pm 50)$ km s$^{-1}$ and the escape velocity $v_{\rm esc}
= (450 \div 650)$ km s$^{-1}$ \cite{Belli:1999nz}. However, the velocity
distribution can be quite different from the simple MB form: it can still be
isotropic but non--maxwellian, as well as non--isotropic. Moreover, the halo
can be affected by some amount of rotation. The way dark matter particles
behave in velocity space affects mostly direct detection, which relies on the
energy transfered from the dark matter particles to a detector.

Proposals of looking for extra--galactic neutralino dark matter have also been
put forward. One possibility is to look for energetic gamma rays from
neutralino annihilation in dense extragalactic systems, like M87 and local
dwarfs spheroidal galaxies \cite{Extra_G:gamma}: in order to have potentially
detectable signals, clumpiness is needed. Another possibility is the direct
detection of dark matter particles coming from outside our Galaxy
\cite{Extra_G:direct}: these particles have a much lower flux than galactic
neutralinos, but they posses some typical features, like an essentially unique
velocity and very few flight directions, which can help in discriminating the
extra--galactic particles from the ones bounded to the Galaxy.

\subsection{Direct detection}
\begin{figure}[t]
\vspace{-31pt}
\includegraphics*[width=0.55\textwidth]{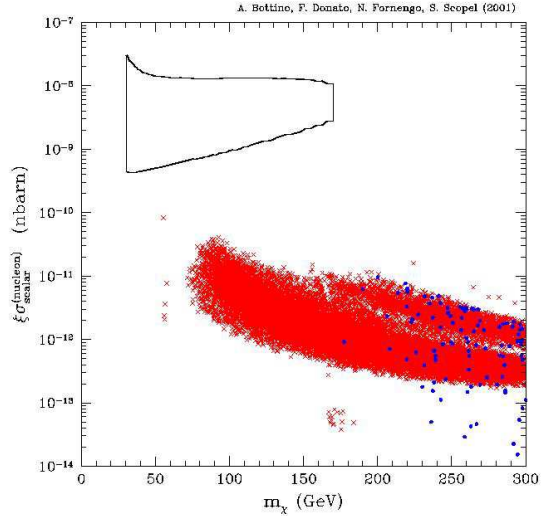}
\vspace{-55pt}
\caption{Relevant direct detection cross section times the fractional
amount of neutralino dark matter in a mSUGRA scheme, with
$m_t$, $m_b$ and $\alpha_s$ fixed at their central values. The (red)
crosses refer to dominant neutralino dark matter, while the (blue)
dots refer to sub--dominant neutralinos ($\Omega_\chi h^2 < 0.05$).}
\label{dsugrastrict}
\end{figure}
\begin{figure}[t]
\vspace{-31pt}
\includegraphics*[width=0.55\textwidth]{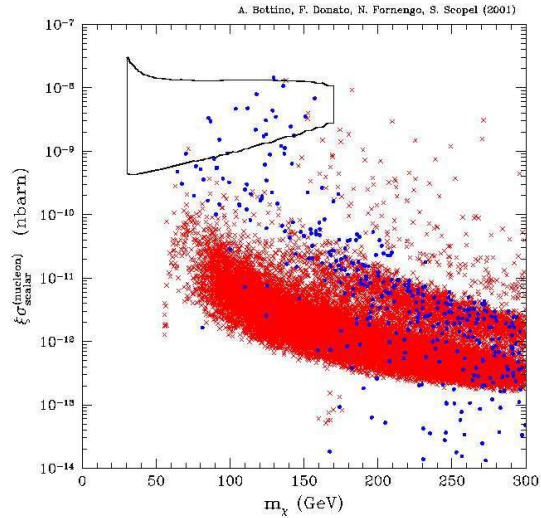}
\vspace{-55pt}
\caption{The same as in fig.\ref{dsugrastrict}, except that $m_t$,
$m_b$ and $\alpha_s$ are varied inside their 2--$\sigma$ allowed
intervals.}
\label{dsugra}
\end{figure}
\begin{figure}[t]
\vspace{-31pt}
\includegraphics*[width=0.55\textwidth]{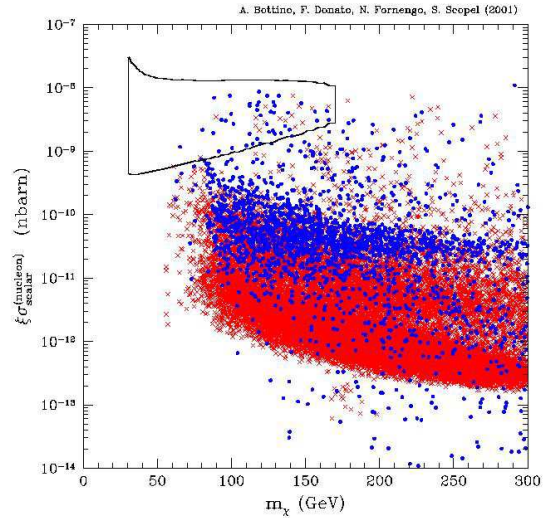}
\vspace{-57pt}
\caption{The same as in fig.\ref{dsugra}, for a 
  nuSUGRA scheme.}
\label{dnusugra}
\end{figure}
\begin{figure}[t]
\vspace{-31pt}
\vspace{13pt}
\includegraphics*[width=0.5\textwidth]{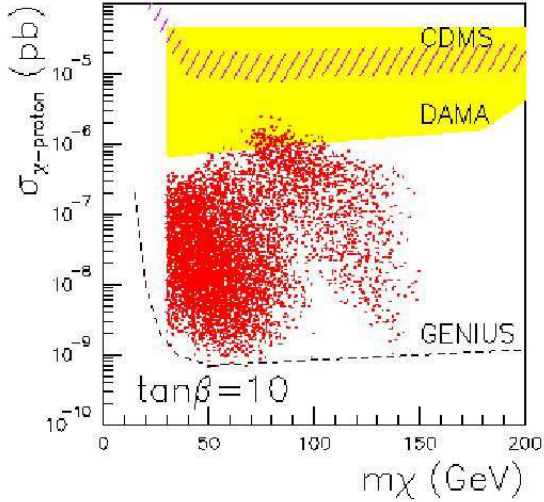}
\vspace{-55pt}
\caption{Relevant direct detection cross section in a D--brane model
with an intermediate unification scale at 10$^{12}$ GeV
\cite{Intermediate-scale.fig}.}
\label{ddbrane}
\end{figure}
\begin{figure}[t]
\vspace{-31pt}
\includegraphics*[width=0.55\textwidth]{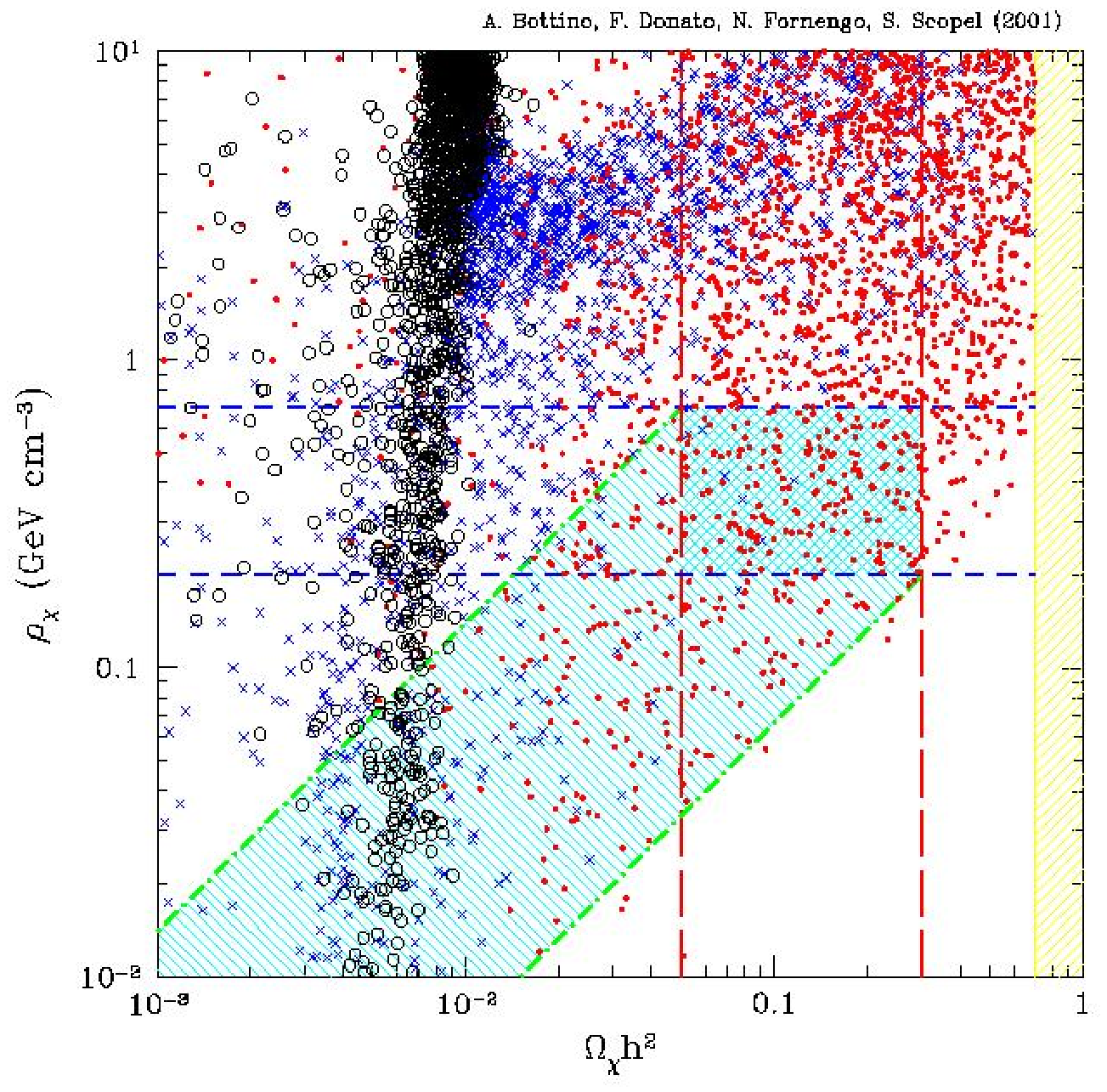}
\vspace{-55pt}
\caption{Neutralino local density $\rho_{\chi}$ derived by requiring
that $[ \rho_{\chi} \sigma_{\rm scalar}^{\rm (nucleon)} ]$ is at the
level of the current experimental sensitivity in direct detection,
plotted vs. the neutralino relic abundance $\Omega_{\chi} h^2$ in the
effMSSM.}
\label{rhoomega}
\end{figure}

Direct detection relies on the scattering of dark matter particles off the
nuclei of a low--background detector. This method is sensitive to the local
properties of the neutralinos in the halo, {\em i.e.} its local abundance
$\rho_\chi$ and its local velocity distribution, and depends on the
neutralino--nucleus scattering cross section, which is usually dominated by the
coherent interaction. The detection rate is proportional to the product $[
\rho_\chi \times \sigma_{\rm scalar}^{\rm (nucleon)} ]$ for any given velocity
distribution \cite{Bottino:2000jx}. The experimental results on direct
detection are reviewed in Ref. \cite{Morales}. We only recall that the current
sensitivity of direct detection experiments can be quoted in the following
ranges for neutralino--{\em nucleon} cross section and mass: $4 \cdot 10^{-9}
{\rm nbarn} \lsim \xi \sigma_{\rm scalar}^{\rm (nucleon)} \lsim {\rm few} \cdot
10^{-8} {\rm nbarn}$ and $30 {\rm ~GeV} \lsim m_\chi \lsim 200 {\rm ~GeV}$
\cite{Bottino:2000jx}.  In the above, we have considered the uncertainties in
the local values of the dark matter density and some uncertainties in the halo
models, discussed in the previous Section. We recall that the quantity $\xi$
measures the fraction of local dark matter to be ascribed to the neutralino
($\xi \leq 1$) \cite{Bottino:2001it}.  This experimental sensitivity of direct
detection is shown in figs.  \ref{dsugrastrict}--\ref{ddbrane} as a closed
region.

The contour which is reported in these figures is actually the DAMA/NaI region
which is obtained when the annual modulation effect observed by the DAMA
Collaboration is interpreted as due to a dark matter particle \cite{DAMA}. The
presence of an annual modulation in the low--energy recoil spectrum is in fact
a distinctive signature of direct detection searches \cite{Freese.mod}.

Figs.\ref{dsugrastrict} and \ref{dsugra} show the comparison between the
experimental results and the theoretical calculations in the mSUGRA scheme. The
quantity $\xi$, which determines whether the neutralino is a dominant or
sub--dominant component, is calculated according to its relic abundance as:
$\xi = {\rm min}(1,\Omega_\chi h^2/0.05)$ \cite{Bottino:2001it}.  The
comparison of the two figures shows that in a constrained mSUGRA
\cite{Berezinsky:1995cj,Bottino:1998ka,Bottino:2000jx,Bottino:2000kq,Ellis.direct-u.omega,Arnowitt.sugra.direct,Vergados.sugra.direct,Drees.sugra.direct,Mandic.direct,Lahanas.sugra.direct,Arnowitt.brane.direct}
scheme the calculation are very sensitive also to standard model parameters
\cite{Bottino:2000kq}: in fig.\ref{dsugrastrict} $m_t$, $m_b$ and $\alpha_s$
are fixed at their own experimental central values, while in fig.\ref{dsugra}
these quantities are allowed to vary inside their 2--$\sigma$ intervals. We
also comment that, especially in strict mSUGRA models, radiative effects are
quite critical, especially in the higgs sector where light higgses ($m_h \sim$
90--100 GeV), not in conflict with the experimental bounds from LEP, can be
obtained for large values of $\tan\beta$ \cite{Bottino:2000kq,Light.Higgs} and
this can induce relatively large values for $\xi \sigma_{\rm scalar}^{\rm
  (nucleon)}$ \cite{Bottino:2000jx,Bottino:2000kq}.  Also the signal--like LEP
higgs events may be compatible with interesting neutralino dark matter
phenomenology for direct detection \cite{Bottino:2000kq}.

Non--universal models
\cite{Berezinsky:1995cj,Bottino:1998ka,Bottino:2000jx,Bottino:2000kq,Ellis.direct-nu.omega,Arnowitt.brane.direct}
are naturally more relaxed and values of $\sigma_{\rm scalar}^{\rm (nucleon)}$
at the level of current sensitivies are more easily obtained, both for a
dominant and a subdominant relic neutralino: an example is shown in
fig.\ref{dnusugra}. Also in D--brane models
\cite{Arnowitt.brane.direct,Nath.brane.direct,Intermediate-scale.fig}, and even
more easily in the effMSSM
\cite{Bottino:2001it,Bottino:2000jx,Bottino:2000kq,Mandic.direct,Bottino:2000gc,Bottino:1999ei,Klapdor.mssm.direct},
cross sections large enough to be probed by direct detection are obtained. Ax
example for a D--brane model with intermediate unification scale is shown in
fig.\ref{ddbrane}.

The question whether current direct detection sensitivities are probing
dominant or subdominant relic neutralinos may be answered in terms of the plot
shown if fig.\ref{rhoomega}, which translates directly in terms of
astrophysical and cosmological quantities the direct detection results
\cite{Bottino:2001it,Bottino:2000jx,Bottino:1999ei,Belli:1999nz,Bottino:1997ru}.
By considering the current interval of sensitivities on the quantity $[
\rho_\chi \times \sigma_{\rm scalar}^{\rm (nucleon)} ]$, the calculation of
$\sigma_{\rm scalar}^{\rm (nucleon)}$ allows us to determine the values of
$\rho_\chi$ which are required for each supersymmetric configuration in order
to provide a detectable signal (for details see for instance Ref.
\cite{Bottino:2000jx} ). Fig.\ref{rhoomega} shows the calculated values of
$\rho_\chi$ vs.  the neutralino relic abundance, for the effMSSM. We see that a
fraction of supersymmetric models overlap with the region of main cosmological
and astrophysical interest: $0.05 \lsim \Omega_\chi h^2 \lsim 0.3$ and $0.2
\leq \rho_\chi/({\rm GeV cm}^{-3}) \leq 0.7$. For points in this region, the
neutralino is the dominant component of dark matter both in the Universe and at
the galactic level. For points which fall inside the band delimited by the
slant dot--dashed lines, the neutralino would provide only a fraction of the
cold dark matter both at the level of local density and at the level of the
average $\Omega$, a situation which would be possible if the neutralino is not
the unique cold dark matter particle component. On the other hand,
configurations above the upper dot--dashed line and below the upper horizontal
solid line would imply a stronger clustering of neutralinos in our halo as
compared to their average distribution in the Universe. This situation may be
considered unlikely, since in this case neutralinos could fulfill the
experimental range for $\rho_\chi$, but they would contribute only a small
fraction to the cosmological cold dark matter content. Finally, configurations
above the upper horizontal line are incompatible with the upper limit on the
local density of dark matter in our Galaxy.

\subsection{Indirect detection}

\begin{figure}[t]
\vspace{-31pt}
\includegraphics*[width=0.55\textwidth]{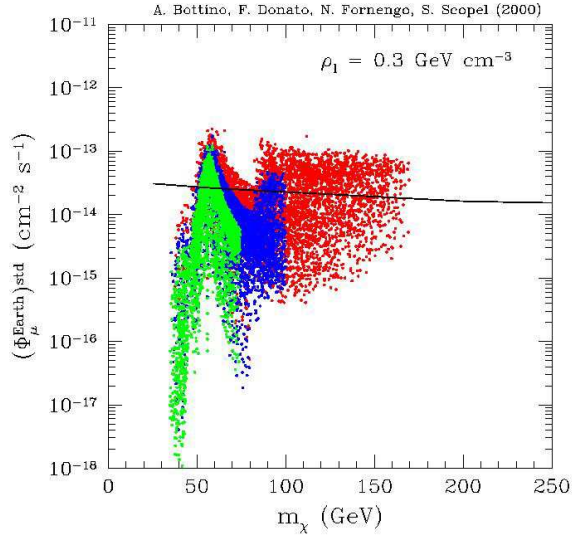}
\label{iearth}
\vspace{-55pt}
\caption{Flux of upgoing muons from neutralino annihilation inside the
  Earth in the effMSSM \cite{Bottino:2000gc}. The horizontal line is the upper
  bound from MACRO.}
\end{figure}
\begin{figure}[t]
\vspace{-31pt}
\includegraphics*[width=0.55\textwidth]{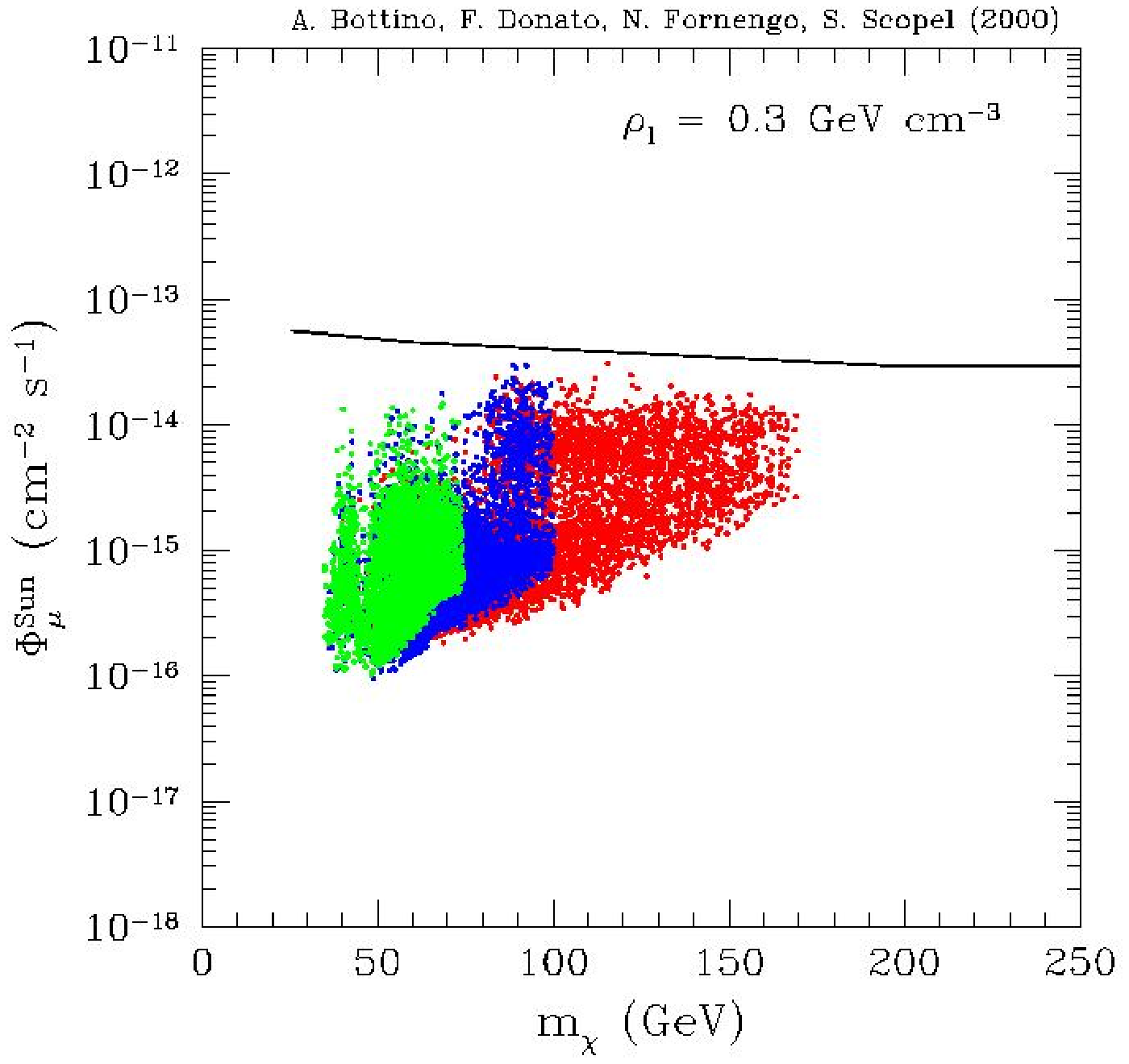}
\label{isun}
\vspace{-55pt}
\caption{Flux of upgoing muons from neutralino annihilation inside the
  Sun in the effMSSM \cite{Bottino:2000gc}. The horizontal line is the upper
  bound from MACRO.}
\end{figure}
\begin{figure}[t]
\vspace{-31pt}
\includegraphics*[width=0.55\textwidth]{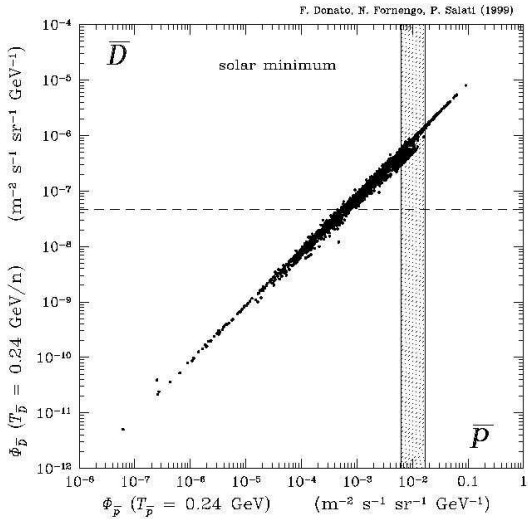}
\label{antimatter}
\vspace{-55pt}
\caption{Flux of antideuterium vs. the flux of antiprotons from
neutralino annihilation in the effMSSM for a smooth halo \cite{Dbar}. The vertical
band is the measurement of BESS95+97. The horizontal line denotes the
reaching capability of AMS on a 3--year flight on board of the space
station}
\end{figure}
\begin{figure}[t]
\vspace{-30pt}
\includegraphics*[width=0.5\textwidth]{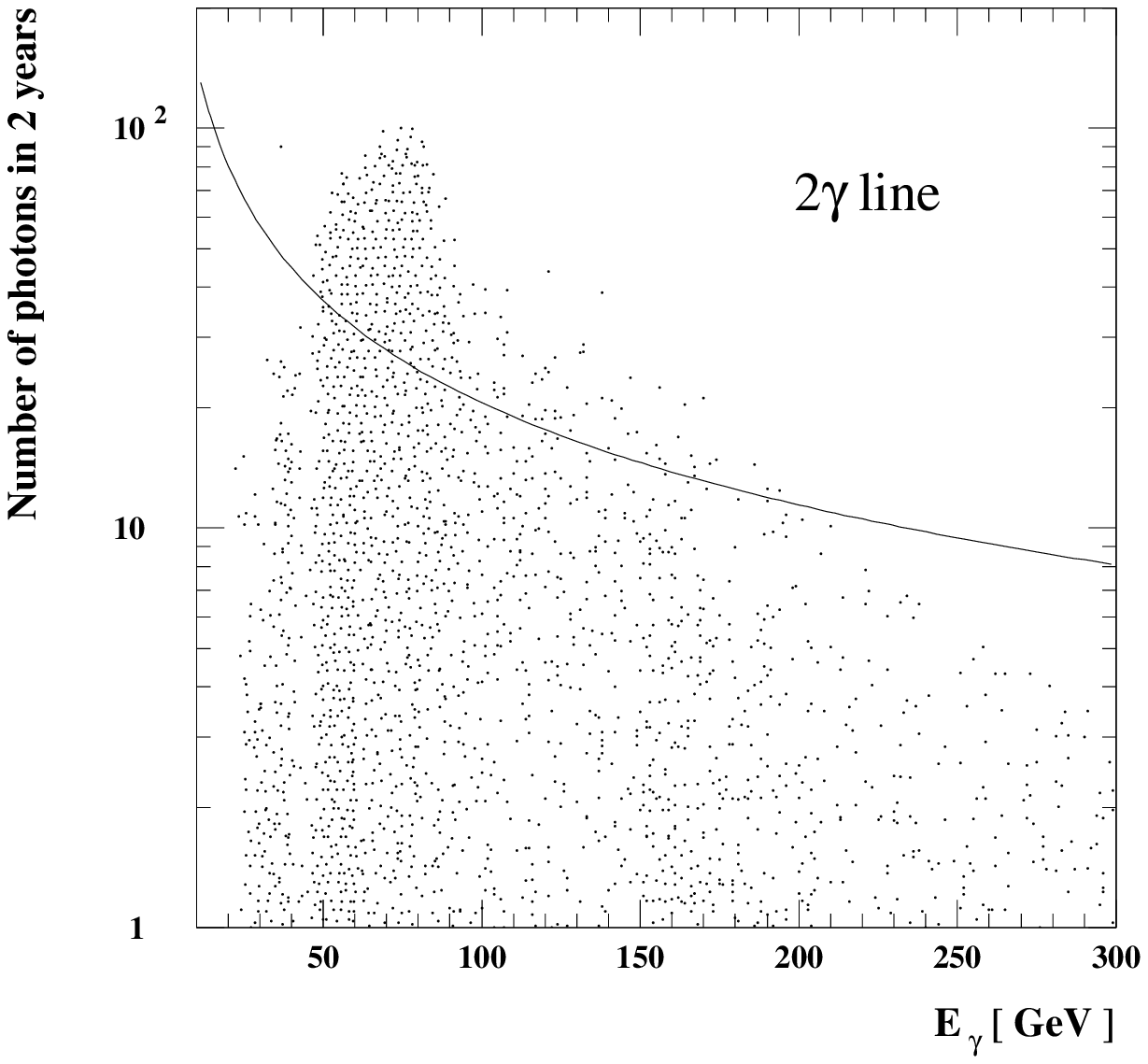}
\vspace{-45pt}
\caption{Predicted gamma--ray line flux in effMSSM for a NFW halo
\cite{Gamma.fig}. The solid line is the estimated GLAST sensitivity.}
\label{gamma}
\end{figure}

Indirect detection relies on the possibility to identify signals which
are originated by neutralino self--annihilations. There are two basic
situations: {\em annihilation inside the Earth or the Sun}, or {\em
annihilation in the galactic halo}.

The first type of signal, due to neutralino annihilation taking place in
celestial bodies (Earth or Sun) where the neutralinos have been gravitationally
captured and accumulated, is a neutrino flux, which can be detected in a
neutrino telescope as a flux of {\em up--going muons}
\cite{Berezinsky:1996ga,Upmu.sugra.u-nu,Bottino:2000gc,Bottino:1998vw,Bottino:1994xp,Bergstrom.mssm.upmu}.
Since the process relevant for accumulation of neutralino is capture, which
relies on neutralino scattering off the nuclei of the Earth and the Sun, this
detection technique is sensitive only to local properties of the Galaxy, like
direct detection. In figs. 7 and 8 we show the flux of upgoing muons calculated
in the effMSSM for the signal coming from the Earth and the Sun, respectively
\cite{Bottino:2000gc}.  The supersymmetric configurations which are shown are
only the ones which are currently at the level of being explored by direct
detection and the upper limit on the upgoing muon flux from MACRO is shown as a
solid line. We can see that indirect detection at neutrino telescopes is
partially competitive with direct detection, although a large fraction of
configurations which are explored by direct detection require more sensitive
neutrino telescopes in order to be probed.

Neutralino annihilations can take place also directly in the galactic halo. In
this case the signals can consist in: a {\em diffuse neutrino flux}
\cite{Clumpy}; a {\em diffuse gamma--ray flux} or a {\em gamma--ray line}
\cite{Gamma,Gamma.fig,Clumpy}; exotic components in cosmic rays: {\em
  positrons} \cite{Clumpy,Positrons.HEAT}, {\em antiprotons}
\cite{Bottino:1998tw,Clumpy,Antiproton} and {\em antideuterium} \cite{Dbar}.
For this type of signals, global properties of the halo are relevant, and
therefore in this case the matter distribution of neutralinos is an important
quantity.  In particular, the overdensities which would be present in a clumpy
halo have the effect of largely enhancing the predicted signals
\cite{Galactic.center,Clumpy}. Neutrino and gamma fluxes require clumpiness to
reach detectable levels. Also the recent anomaly in the HEAT data \cite{HEAT}
on positrons may be explained by a signal originated by neutralino annihilation
if overdensities are present in the galactic halo \cite{Positrons.HEAT}. On the
contrary, antiprotons and antideuterium may be detectable also in a smooth
halo.

In fig. 9 we show the correlation between the antiproton and the antideuterium
flux calculated in the effMSSM for a smooth halo \cite{Dbar}. The vertical band
represents the BESS 95+97 measurement of the antiproton flux. All the points on
the right of this band have to be considered as excluded, since they provide a
flux in excess of the measurements: this fact shows that the antiproton signal
can be used to constrain the supersymmetric parameter space.  At present, there
are no measurement of an antideuterium component in cosmic rays, but the
analysis of Ref. \cite{Dbar} shows that the antideuterium signature is quite
promising and the next generation detectors may be able to probe a sizeable
fraction of supersymmetric configurations. This is shown in fig. 9 by the
horizontal line, which denotes the reaching capability of the AMS detector on a
3--year flight on board of the space station. Recently, a detector designed
specifically to look for antideuterium in cosmic rays has been proposed
\cite{Dbar.detector}.

Finally, fig. \ref{gamma} shows the predicted gamma--ray line flux in the
effMSSM for a NFW halo, compared with an estimation of the GLAST sensitivity.

\section{Conclusions}
We have seen that we can identify two main issues in particle dark matter
studies: {\em i)} to explain the observed amount of dark matter in the Universe
($0.05 \lsim \Omega_{\rm M} h^2 \lsim 0.3$) by finding suitable particle
candidates; {\em ii)} to detect a relic particle. For both of them there appear
to be good prospects of success.

As for the candidates, there are many proposed particles which could act as
dark matter. Some of these candidates turn out to be quite natural, like {\em
  e.g.} the massive neutrino, the axion or the neutralino. Almost all of the
proposed candidates can play the role of the dominant dark matter component,
although for some of them a non--standard cosmology is required. An important
remark is that, from the particle physics point of view, dark matter may
naturally be multi--component. A multi--component dark matter scenario offers
opportunity for interesting phenomenology not only to the dominant candidate,
which would explain the cosmological observation on the $\Omega_{\rm M}$
parameter, but also to the sub--dominant candidates, since usually these are
the ones which are easier to detect. The detection of a particle which is a
relic from the early Universe would be a very important and exciting result.

As for detection, perspectives are good, both for direct and for indirect
detection techniques, especially for the most interesting and studied
candidate, the neutralino. The possibility to have detectable rates for
neutralinos depends on the specific supersymmetric model which is considered,
and quite generally it appears simpler to detect a relic neutralino which is a
sub--dominant dark matter component.  Nevertheless, there are many
supersymmetric schemes where relic neutralinos can provide enough cosmological
abundance to explain the observed amount of dark matter, and at the same time
they can have detection rates large enough to be accessible to direct, and also
to some indirect, detection methods. The positive indication of annual
modulation in the detection rate of the DAMA/NaI Collaboration can be
interpreted as pointing towards a direct detection of a relic massive particle.
This effect is, at the moment, the most compelling indication for a particle
dark matter signal. When interpreted as originated from relic neutralinos, the
annual modulation effect can be explained in a number of realization of
supersymmetry. It is worth noticing that the presence of a signal from dark
matter, like the annual modulation effect or signals which could hopefully come
in future experiments, can be very important not only for astrophysics and
cosmology but also for particle physics, since the need to explain the effect
can help in deriving properties of particle physics models and possibly
discriminate among different realizations, for instance of supersymmetry.

\smallskip
{\bf Acknowledgements.} This work was partially supported by the
Research Grants of the Italian Ministero dell'Universit\`a e della
Ricerca Scientifica e Tecnologica (MURST) within the {\sl
Astroparticle Physics Project}. I also wish to thank the Organizers
for inviting me to deliver this talk at the TAUP 2001 Conference.

 \end{document}